\documentclass[reprint,amsmath,amssymb,aps,pre]{revtex4-2}
\usepackage{graphicx}
\usepackage{dcolumn}
\usepackage{bm}
\usepackage{amsmath}
\usepackage{booktabs}

\begin{document}

\title{Disordered hyperuniform modulated phases and the cosmic web from one free energy}

\author{Fausto Martelli}
\email{fausto.martelli@cnr.it}
\affiliation{CNR-Istituto dei Sistemi Complessi, P.le Aldo Moro 5, 00185 Rome (Italy)}
\affiliation{Dipartimento di Fisica, Universit\'a La Sapienza di Roma, P.le Aldo Moro 5, 00185 Rome (Italy)}

\date{\today}

\begin{abstract}
Disordered hyperuniformity spans twenty-five orders of magnitude, from the
microscopic world of soft and condensed matter to the distribution of matter
in the universe. We introduce a free-energy functional with no adjustable
coupling, a phase-field-crystal ordering term and a Newtonian tail, and show
that its two kinetic limits generate two of these structures. Under conserved
overdamped dynamics it arrests in a bicontinuous labyrinth of the kind found
in confined fluids, block copolymers, active matter and vegetation patterns,
which is a class III disordered hyperuniform state; under inertial dynamics in
an expanding background it generates a cosmic web whose tidal skeleton matches
that of gravity alone to within $0.03$ in every morphological class. In both
limits the suppressed long-wavelength fluctuations are inherited from the
initial condition rather than made by the dynamics: the large-scale exponent
of the arrested state equals the primordial one, and gravity rescales the
infrared spectrum by the linear growth factor to better than one per cent over
a factor $1363$ of growth. The two sectors are continuously connected, and
their distance is measurable: once the ordering modulation saturates, its
mean-field back-reaction leaves the medium with a residual pressure
proportional to $\lambda_0^2$ that suppresses the growth of the largest scales
by a smooth factor and vanishes in the cold-matter limit $\lambda_0\to0$.
\end{abstract}

\maketitle

\section{Introduction}
Disordered hyperuniformity (DHU) denotes the anomalous suppression of long-wavelength density fluctuations, $S(k)\rightarrow0$ as $k\rightarrow0$, in systems that possess no crystalline order~\cite{torquato2003local}. DHU is ubiquitous and has been identified in a wide range of systems spanning multiple length scales, including amorphous materials~\cite{hejna2013nearly,xie2013hyperuniformity,martelli2017large,zheng2020disordered,martelli2022steady,chen2023disordered,formanek2023molecular} and superconductors~\cite{le2017enhanced,llorens2020disordered,rumi2019hyperuniform}, metamaterials~\cite{tang2022soft,gallego2025hole,maher2025characterizing}, confined fluids~\cite{leoni2025confinement}, jammed packings~\cite{atkinson2014existence,klatt2014characterization,tian2015geometric,wang2025hyperuniform}, suspensions~\cite{hexner2017enhanced}, polymer melts~\cite{chen2024emergence}, cellular  rganizations~\cite{wax2002cellular,klatt2019universal}, avian photoreceptors~\cite{jiao2014avian}, leaf vein  networks~\cite{liu2024universal}, active matter~\cite{huang2021circular}, vegetation patterns in drylands~\cite{hu2025causes}, the collective movement of large animal groups~\cite{lu2026orientation}, the organization of pebbles on sand on Mars and Earth~\cite{zhu2026diverse}, and the large-scale structure of the universe~\cite{pietronero2002statistical,gabrielli2002glass,pietronero2005statistical,gabrielli2005statistical,philcox2023disordered}. Such ubiquity indicates that DHU is a common feature shared by systems that are otherwise treated and studied as independent entities.

Here we ask whether such diverse systems can be described by one free energy under different kinetic limits. We introduce a functional with two competing ingredients: a finite-wavelength ordering instability at $k_0$, which drives a uniform system towards a solid-like state, and a Newtonian tail, which amplifies the longest wavelengths below the Jeans wavenumber $k_J$. Under conserved overdamped dynamics the functional arrests in a DHU labyrinth, an arrangement that regulates density fluctuations on multiple length- scales~\cite{merminod2015transition,ma2017random,huang2021circular,leoni2025confinement}; under inertial dynamics in an expanding background it generates a cosmic web. The fluid fixes the gravitational coupling.

The paper is organised as follows. Section~\ref{sec:model} constructs the functional from classical density functional theory, derives its linear stability and shows that its two kinetic limits descend from one kinetic equation. Section~\ref{sec:methods} describes the initial fields and the numerical schemes. Section~\ref{sec:soft} presents the conserved limit: the survival of the infrared spectrum through a quench, its amplification by gravity, and the real-space identification of the arrested state as a class III DHU labyrinth. Section~\ref{sec:cosmo} presents the inertial limit: the cosmic web under the full functional, the crossover to gravity as $\lambda_0$ is lowered, a paired measurement of what the ordering term does to the infrared, and the quantitative infrared test in the cold-matter limit. Section~\ref{sec:conc} reports our final discussions and conclusions.

\section{The free energy and its two limits}
\label{sec:model}

\subsection{The free energy}
We describe matter by the dimensionless density contrast $n(\mathbf{x})=\delta\rho/\bar{\rho}$, where $\bar{\rho}$ is the mean density. The free energy is
\begin{equation}
 \begin{split}
 F[n]=&\int d^3x\left[\frac{n}{2}\left(r+\left(1+\nabla^2\right)^2\right)n+\frac{n^4}{4}\right]\\
 &-\frac{G}{2}\int\frac{d^3k}{(2\pi)^3}\frac{|\tilde{n}(\mathbf{k})|^2}{k^2}
 \end{split}
 \label{eq:free}
\end{equation}
where $\tilde{n}(\mathbf{k})=\int d^3x\,e^{-\imath\mathbf{k}\cdot\mathbf{x}}n(\mathbf{x})\equiv n_k$. The first term in eq.~\ref{eq:free} is a canonical phase-field-crystal free energy~\cite{elder2004modeling,elder2007phase} in which the parameter $r$ controls the ordering transition, while $(1+\nabla^2)^2$ vanishes at the wavenumber $k_0\equiv1$ in the adopted units. The quartic term saturates the growth of unstable fluctuations at finite amplitude and endows $F$ with metastable minima~\cite{elder2004modeling,elder2007phase}. The last term in eq.~\ref{eq:free} is Newtonian gravity, with $G$ a dimensionless gravitational coupling defined below.

Equation~\ref{eq:free} follows from classical density functional theory. To quadratic order, the free energy of a uniform fluid can be written as $\beta F_2=\frac{\bar{\rho}}{2}\int\frac{d^3k}{(2\pi)^3}\left[1-\bar{\rho}\hat{c}_2(k)\right]|\tilde{n}(\mathbf{k})|^2$, where $\hat{c}_2(k)$ is the Fourier transform of the direct correlation function. We decompose the pair potential into a short-ranged contribution and a Newtonian tail, $-G_Nm^2/r$, with $G_N$ the gravitational constant and $m$ the particle mass. Expanding the short-range contribution to $\hat{c}_2(k)$ about its first maximum~\cite{elder2007phase} and treating the Newtonian tail within the random-phase approximation, which becomes asymptotically exact for long-range interactions, yields eq.~\ref{eq:free}. The resulting dimensionless gravitational coupling is
\begin{equation}
    G=\frac{4\pi\beta G_Nm^2\bar{\rho}}{B_xk_0^2},
    \qquad B_x=-\frac{1}{8}k_0^2\bar{\rho}\,\hat{c}_2^{\,\prime\prime}(k_0)
    \label{eq:G}
\end{equation}
with $\beta$ the inverse thermal energy. Once the fluid and its mean density are specified, $G$ is fixed by the microscopic interactions. Likewise $\lambda_0=2\pi/k_0$ is set by the first peak of $\hat{c}_2$ and is a property of the medium.

\subsection{Linear stability}
\label{sec:stability}
\begin{figure}[t]
\includegraphics[width=\columnwidth]{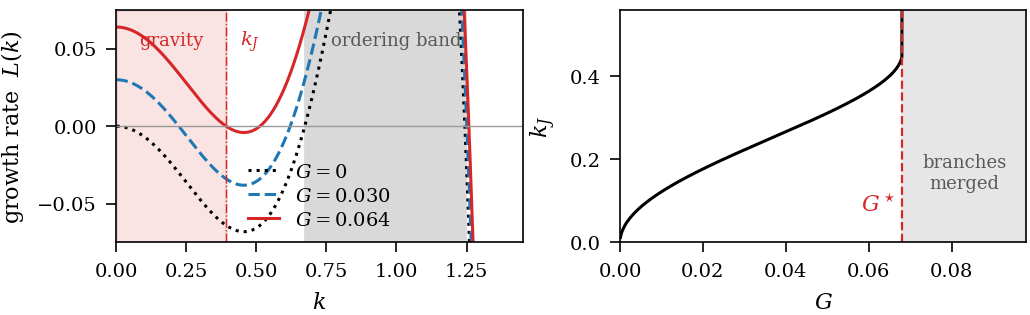}
\caption{Left: growth rate $L(k)$, eq.~\ref{eq:LL}, at $r=-0.3$. The red
shaded area is the gravitational band $k<k_J$, the grey shaded area the
ordering band around $k_0$. Right: the Jeans branch $k_J(G)$ terminates at
$G^{\star}=0.0681$, eq.~\ref{eq:jeans}.}
\label{fig:model}
\end{figure}

A small perturbation $n_ke^{L(k)t}$ of the uniform state grows at the rate
\begin{equation}
    L(k)=-k^2\left[r+(1-k^2)^2\right]+G .
    \label{eq:LL}
\end{equation}
The case $r<0$ generates a short-range ordering band $(1-k^2)^2<-r$ around $k_0$ (grey shaded area in fig.~\ref{fig:model}, left) and, separately, $L(k)$ is positive for all $k<k_J$, where the Jeans wavenumber $k_J$ is, with $u=k^2$, the smallest positive root of
\begin{equation}
    u^3-2u^2+(1+r)u=G .
    \label{eq:cubic}
\end{equation}
For $k\ll k_0$ the growth rate reduces to $L(k)\simeq G-(1+r)k^2$, so the marginal mode gives $k_J^2=G/(1+r)$~\footnote{Restoring dimensions, using $G=4\pi\beta G_Nm^2\bar{\rho}/(k_0^2v_{th}^2)$ and $c_s^2=v_{th}^2(1+r)$, yields the standard Jeans criterion $k_{J,\rm standard}^2=4\pi G_N\bar{\rho}/c_s^2$.}. The limit $k\rightarrow0$ of the short-range kernel, $r+1$, therefore plays the role of an inverse compressibility.

At $r=-0.3$ the left-hand side of eq.~\ref{eq:cubic} has a local maximum at $u^{*}=0.207$. The gravitational branch (red shaded area in fig.~\ref{fig:model}, left) therefore exists only for $G<G^*=0.0681$ and, at the bifurcation, $k_J=0.455k_0$. In terms of the Jeans length $\lambda_J=2\pi/k_J$, a kinetically arrested disordered state can exist only where
\begin{equation}
    \lambda_J>2.20\lambda_0 .
    \label{eq:jeans}
\end{equation}
For $G>G^*$ the gravitationally unstable range overlaps the ordering band, preventing a kinetically arrested state (fig.~\ref{fig:model}, right). Equation~\ref{eq:jeans} contains no adjustable parameters.

In the cosmological sector the medium is pressureless. Since $r+1$ is the inverse compressibility, cold matter corresponds, at the linear level, to the point $r=-1$, with $\lambda_0$ below the structure-forming scales; the nonlinear correction to this statement is measured in Sec.~\ref{sec:pair}. Equation~\ref{eq:jeans} is specific to the value of $r$ at which it is evaluated and characterises the arrested sector. At $r=-1$ the coefficient $(1+r)$ vanishes, eq.~\ref{eq:cubic} reduces to $u^3-2u^2=G$, and its derivative $3u^2-4u$ vanishes only at $u=4/3$, where the left-hand side is negative: there is no turning point at positive $G$, hence no $G^*$, and the Jeans branch does not terminate. Its smallest positive root lies at $u\simeq2.02$ for $G=0.065$, that is
\begin{equation}
    k_J=1.42\,k_0,\qquad\lambda_J=0.70\,\lambda_0 ,
\end{equation}
so the gravitationally unstable range extends past the ordering wavelength for every $G$ and the two instabilities are permanently merged. This does not violate eq.~\ref{eq:jeans} but complements it: the same criterion that permits a kinetically arrested state at $r\simeq-0.3$ forbids one in the pressureless limit, which is why the inertial runs of Sec.~\ref{sec:cosmo} produce a cosmic web rather than an arrest.

\subsection{Two kinetic limits from one kinetic equation}
\label{sec:limits}
The overdamped and inertial regimes, corresponding to the soft-matter and the cosmological limit respectively, are governed by the same chemical potential
\begin{equation}
    \mu_k\equiv\left.\frac{\delta F}{\delta n}\right|_k
    =\left[r+(1-k^2)^2\right]n_k+(n^3)_k-\frac{G}{k^2}n_k
    \label{eq:mu}
\end{equation}
and differ by how density is transported. Both are limits of one kinetic equation. Consider $N$ particles of mass $m$ interacting via the potential $v$, in a bath at temperature $T$, with friction $\zeta$. In the mean-field description the one-particle distribution $f(\mathbf{x},\mathbf{u},t)$ obeys the Kramers equation~\cite{risken1989fokker}
\begin{equation}
    \partial_tf+\mathbf{u}\cdot\nabla f-\frac{\nabla\mu}{m}\cdot\partial_{\mathbf{u}}f
    =\zeta\,\partial_{\mathbf{u}}\cdot\left(\mathbf{u}f+\frac{1}{\beta m}\partial_{\mathbf{u}}f\right),
    \label{eq:kramers}
\end{equation}
where $\mu=\delta F/\delta n$ and $n(\mathbf{x},t)=\int d^3u\,f$. In the limit $N\rightarrow\infty$ at fixed $G_NmN$ the mean-field closure is exact~\cite{braun1977vlasov}, and the potential entering $\mu$ is that of eq.~\ref{eq:free}.

In the overdamped limit $\zeta\rightarrow\infty$, expanding eq.~\ref{eq:kramers} in $1/\zeta$, the velocity distribution is Maxwellian on timescales of order $\zeta^{-1}$ and the density follows the Smoluchowski equation~\cite{marconi1999dynamic}, $\partial_tn=(m\zeta)^{-1}\nabla\cdot(n\nabla\mu)$, with mobility $M=n/(m\zeta)$. Linearising the mobility and setting it to unity, density is conserved and inertia is negligible:
\begin{equation}
    \partial_tn_k=-k^2\mu_k ,
    \label{eq:cons1}
\end{equation}
with zero thermal noise. The factor $k^2$ is the signature of conserved dynamics: mass must be transported over a distance $k^{-1}$, so relaxation slows as $k^{-2}$ and the longest wavelengths freeze.

In the frictionless limit $\zeta\rightarrow0$, eq.~\ref{eq:kramers} reduces to the Vlasov equation, whose first two velocity moments are $\partial_tn+\nabla\cdot(n\mathbf{v})=0$ and $\partial_t\mathbf{v}+(\mathbf{v}\cdot\nabla)\mathbf{v}=-\nabla\mu/m$. Transforming to comoving coordinates and linearising in the velocity gives the cosmological limit, in which inertia is retained and peculiar velocities are damped by the expansion:
\begin{equation}
    \begin{split}
        &\partial_tn=-\tilde{\theta},\\
        &\partial_t\tilde{\theta}=-2H\tilde{\theta}+k^2\mu_k ,
    \end{split}
    \label{eq:cons}
\end{equation}
where $a$ is the scale factor, $\tilde{\theta}=\nabla\cdot\mathbf{v}/a$ is the scaled velocity divergence and $H=\dot a/a$ is the Hubble rate. We adopt a $\Lambda$CDM background, $H(a)=H_0(\Omega_ma^{-3}+\Omega_\Lambda)^{1/2}$, with $H_0=H(a=1)$ the present-day Hubble parameter and $\Omega_m=0.3153$, $\Omega_\Lambda=1-\Omega_m=0.6847$ the present-day density parameters of matter and of the cosmological constant~\cite{Planck2020}. At sufficiently small $k$, eq.~\ref{eq:cons} reduces to the standard linear gravitational growth equation. Friction is thus the single parameter that interpolates between the two limits.

The longest wavelengths are frozen in the conserved limit but grow scale-independently in the cosmological one. In both, the question is whether the nonlinear dynamics generates a universal infrared spectrum or preserves the one already present in the initial field. We distinguish these possibilities by evolving fields with identical short-scale structure but different infrared spectra.

\section{Initial fields and numerical methods}
\label{sec:methods}

\subsection{Initial fields}
\begin{figure}[t]
\includegraphics[width=\columnwidth]{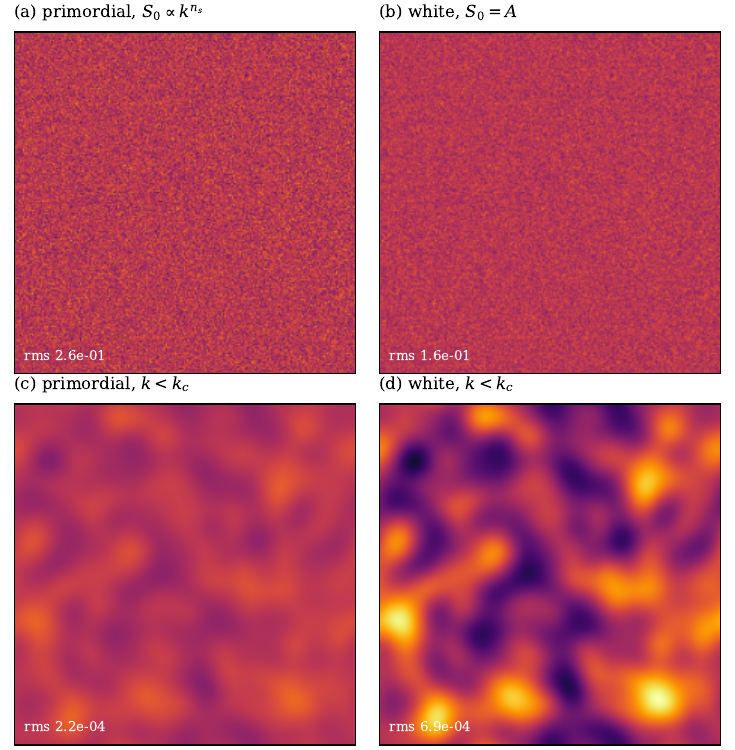}
\caption{Panels (a) and (b): primordial scalar field with $S_0(k)\propto
k^{n_s}$ and white-noise field with $S_0(k)=A$. Panels (c) and (d): the same
fields low-pass filtered with $\exp[-(k/k_c)^2/2]$, which retains $k<k_c$.
Plots share the same colour scale within each row. Rendered raw, (a) and (b)
are indistinguishable, which is the design: it rules out a difference in
short-scale structure as the origin of the spectral separation measured in
Sec.~\ref{sec:soft}.}
\label{fig:initial}
\end{figure}

We characterise the spectrum through the structure factor $S(k)=\langle|n_{\mathbf{k}}|^2\rangle_{|\mathbf{k}|=k}$, where the average is taken over spherical Fourier shells. The primordial field is initialised with
\begin{equation}
    S_0(k)=Ak^{n_s},
    \label{eq:primordial}
\end{equation} 
where $n_s=0.9665(\pm0.0038)$ is the scalar spectral index inferred from CMB observations~\cite{Planck2020}. As a control, we construct a second field with the same short-scale realisation but a white-noise infrared spectrum, $S_0(k)=A$. The two fields are built from the same random realisation and normalised to the same amplitude in the ordering band, so they are statistically indistinguishable at short wavelengths; the white field carries $3.2$ times more power below the freezing scale $k_c$ defined in Sec.~\ref{sec:soft}. Any difference in the final spectra therefore originates at $k<k_c$ and cannot be attributed to a difference in small-scale structure (fig.~\ref{fig:initial}).

\subsection{Conserved overdamped dynamics}
Equation~\ref{eq:cons1} is integrated pseudo-spectrally on a periodic cubic grid with $5.5$ collocation points per ordering wavelength $\lambda_0$, so that the ordering band is resolved with margin against the grid Nyquist frequency. The linear operator is treated semi-implicitly and the cubic term explicitly, evaluated in real space and transformed back at each step. The mean density is re-projected onto its target value at every step; without this, accumulated round-off in the $k=0$ coefficient displaces $\bar n$ over the $\sim10^4$ steps of a quench. Time is measured in natural units of the conserved dynamics, obtained by setting the mobility in  eq.~\ref{eq:cons1} to unity. Parameters are collected in Table~\ref{tab:soft}.

\begin{table*}[b]
\caption{Parameters of the conserved overdamped (soft-matter) runs.}
\label{tab:soft}
\begin{ruledtabular}
\begin{tabular}{lll}
Quantity & Symbol & Value \\ \colrule
Box size                   & $L_{\rm box}$ & $48\lambda_0$, $96\lambda_0$ \\
Grid points per wavelength &               & $5.5$ \\
Mean density offset        & $\bar n$      & $-0.25$ ($-0.34$ for the $\bar n^2$ test) \\
Effective temperature ramp & $r_{\rm eff}$ & $0.4\rightarrow-0.3$ over $\tau/2$ \\
Total time                 & $\tau$        & $120$ \\
Gravitational coupling     & $G$           & $0$ and $0.01$ \\
Freezing scale             & $k_c$         & $\simeq0.08\,k_0$ \\
Realisations               &               & 3 per box %% CHECK
\end{tabular}
\end{ruledtabular}
\end{table*}

\subsection{Inertial dynamics in an expanding background}
\label{sec:pm}
Equation~\ref{eq:cons} is integrated in Lagrangian form with a particle-mesh scheme, so that matter advection is represented explicitly. Density is assigned by cloud-in-cell deposition; the force is obtained in Fourier space as $\mathbf{F}(\mathbf{k})=-i\mathbf{k}\,\tilde v(k)\,\tilde n(\mathbf{k})$ with
\begin{equation}
    \tilde v(k)=\frac{4\pi G_N\bar\rho}{Gk_0^2}
    \left[\,r+\left(1-u^2\right)^2-\frac{G}{u^2}\right],
    \qquad u=\frac{k}{k_0}
    \label{eq:kernel}
\end{equation}
so that the $-G/u^2$ term is identically $-4\pi G_N\bar\rho/k^2$ and the ordering bracket can be switched off exactly, without altering the gravitational sector; we refer to this as the $\lambda_0\rightarrow0$ limit. The cubic term of eq.~\ref{eq:mu} is evaluated in real space and added to the potential with the same normalisation. Trajectories are advanced with a second-order symplectic kick--drift--kick leapfrog using $60$ steps uniform in $\ln a$ from $a=0.02$ ($z=49$) to $a=1$, and initial displacements are Zel'dovich displacements of a regular lattice. The background is that of Sec.~\ref{sec:limits} with $h=0.6736$, and the initial field is normalised so that the linearly extrapolated $\sigma_8=0.8111$ at $a=1$~\cite{Planck2020}.

The ordering band extends to $1.24\,k_0$ and must lie below the force cutoff
at $0.6\,k_{\rm Nyq}$, which sets a resolution bound
\begin{equation}
    N_{\rm mesh}>\frac{2.07\,k_0L_{\rm box}}{\pi}.
    \label{eq:resbound}
\end{equation}
Undersizing the mesh removes the ordering instability rather than producing a
visible failure, so eq.~\ref{eq:resbound} is checked at run time for every
configuration in Table~\ref{tab:cosmo}.

Cosmic-web morphology is classified with the tidal tensor
$T_{ij}=\partial_i\partial_j\Phi$, evaluated from the density smoothed on
$R_s$, by counting eigenvalues exceeding $\lambda_{\rm th}=0.2$, the number of
directions along which matter is collapsing. The smoothing radius is set to
two mesh cells with a floor at $2\,h^{-1}$Mpc, and comparisons between runs
are made only at equal $R_s$.

\begin{table*}[b]
\caption{Parameters of the inertial (cosmological) runs. All start from
$a=0.02$ and end at $a=1$ with $60$ steps uniform in $\ln a$; $\lambda_0\to0$
denotes the limit in which the ordering bracket of eq.~\ref{eq:kernel} is
inactive at every resolved scale. The two ``paired'' runs share the initial
field and the mesh.}
\label{tab:cosmo}
\begin{ruledtabular}
\begin{tabular}{lccccccc}
Run & $L_{\rm box}$ & $N_{\rm mesh}$ & $N_{\rm part}$ & cell & $k_0$ & $r$ & $G$ \\
    & [$h^{-1}$Mpc] & & & [$h^{-1}$Mpc] & [$h\,$Mpc$^{-1}$] & & \\ \colrule
$\lambda_0\to0$, 3 seeds & 1200 & $400^3$ & $400^3$ & 3.00  & --- & --- & --- \\
$\lambda_0\to0$, 4 seeds & 300  & $128^3$ & $128^3$ & 2.34  & --- & --- & --- \\
full functional, paired  & 300  & $800^3$ & $800^3$ & 0.375 & 4.0 & $-1$ & 0.065 \\
$\lambda_0\to0$, paired  & 300  & $800^3$ & $800^3$ & 0.375 & --- & --- & --- \\
full functional          & 150  & $400^3$ & $400^3$ & 0.375 & 4.0 & $-1$ & 0.065 \\
full functional          & 300  & $400^3$ & $400^3$ & 0.75  & 2.0 & $-1$ & 0.065 \\
full functional          & 1200 & $400^3$ & $400^3$ & 3.00  & 0.5 & $-0.4$ & 0.050 \\
full functional          & 1200 & $400^3$ & $400^3$ & 3.00  & 0.5 & $-0.3$ & 0.030 \\
\end{tabular}
\end{ruledtabular}
\end{table*}

\section{The conserved limit: a hyperuniform labyrinth}
\label{sec:soft}

\subsection{Spectral survival and inheritance}
\begin{figure*}[t]
\includegraphics[width=\textwidth]{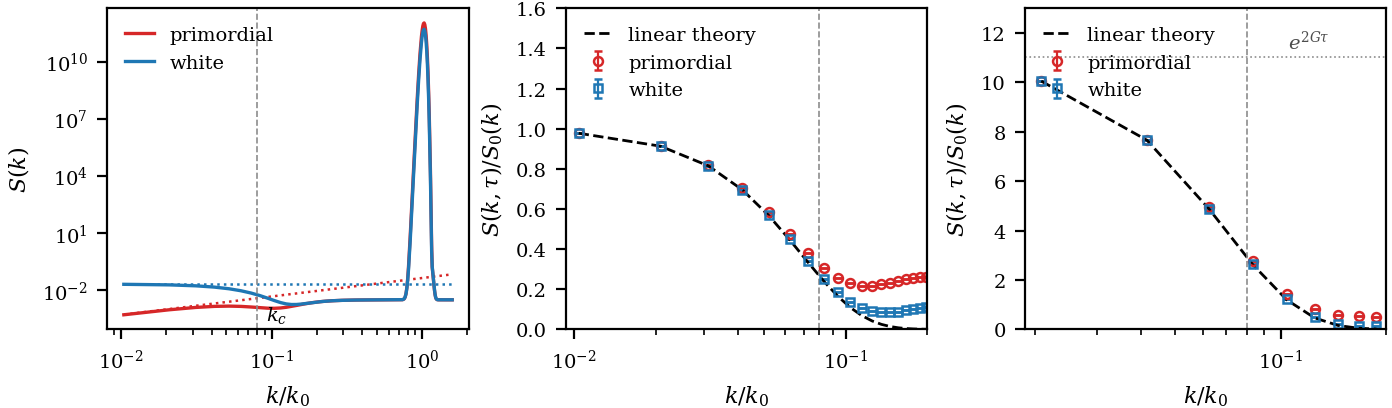}
\caption{Left: initial (dotted) and final (solid) spectra of the
primordial-spectrum and white-noise fields for $L_{\rm box}=96\lambda_0$.
Below $k_c$ each final spectrum tracks its own initial condition, while above
$k_c$ both converge on the dynamics-generated spectrum; the peak at $k_0$
signals the kinetically arrested amorphous state. Middle: mode-survival ratio
against the parameter-free linear prediction. Right: as in the middle panel,
with the gravitational term switched on ($G=0.01$); the same shells are
amplified $e^{2G\tau}=11.0$-fold, shape-preserved.}
\label{fig:inherit}
\end{figure*}

We first explore the soft-matter limit of eq.~\ref{eq:free}, simulating quenching runs with no gravity term ($G=0$) in the overdamped conserved dynamics eq.~\ref{eq:cons1} at a mean density offset $\bar n=-0.25$. This choice is justified by noticing that introducing $n=\bar n+\varphi$~\footnote{The quantity $\varphi$ is the fluctuation of the density contrast about its mean: $\varphi(\mathbf{x})\equiv n(\mathbf{x})-\bar n$.} in eq.~\ref{eq:free} generates a shift of the effective temperature $r_{\rm eff}=r+3\bar n^2$ and a three-wave coupling $3\bar n\varphi^2$, the only channel that allows short-wavelength modes to feed the large scales. We ramp $r_{\rm eff}$ from $0.4$ to $-0.3$ over the first half of a total time $\tau=120$, in boxes of $L_{\rm box}=48\lambda_0$ and $96\lambda_0$, three independent realisations each. The conserved dynamics relaxes a mode of wavenumber $k$ at a rate proportional to $k^2$; setting $2(1+r)k^2\tau\simeq1$ locates the boundary $k_c\simeq0.08$ below which modes are unaffected by the quench (fig.~\ref{fig:inherit}, left).

From linear theory, the surviving fraction of each mode is
$S(k,\tau)/S_0(k)=\exp\left[2\int_0^\tau L(k,t)\,dt\right]$, with $L$ of
eq.~\ref{eq:LL} evaluated along the ramp: a parameter-free prediction. Below
$k_c$ the measured survival of both fields follows it with no fitted
parameters (fig.~\ref{fig:inherit}, middle): the ratio falls from unity at the
largest scales to $\simeq0.25$ at $k_c$, tracking the linear curve to within
the shell-to-shell scatter. Above $k_c$ both fields exceed it, by an amount
discussed below.

At small $k$, $97\%$ of the initial power is preserved at the end of the
quench, and the primordial field and the white control agree with each other
to $0.5\%$ (the overlap of dotted and solid lines in fig.~\ref{fig:inherit},
left), indicating that the dynamics acts identically on the two spectra;
convergence between them occurs only above $k_c$. The large-scale exponent
$\alpha\equiv d\ln S/d\ln k$ in the box $L_{\rm box}=48\lambda_0$ is
$\alpha=0.93\pm0.20$ for the primordial field, against $n_s=0.9665$, and
$\alpha=-0.43\pm0.46$ for the control. Their separation, $1.36\pm0.50$, is
consistent with the $n_s$ that separated them initially. The large-scale exponent of the arrested state is therefore inherited from the initial condition rather than generated by the dynamics.

\subsection{Mode coupling above $k_c$}
For $k>k_c$ the measured power exceeds linear theory. Linear theory cannot transfer power between scales, so the excess arises from mode coupling. At lowest order, the term $3\bar n\varphi^2$ combines two short-wavelength modes into a long-wavelength one, with a contribution scaling as $\bar n^2$. Repeating the quench at $\bar n=-0.34$ increases the excess by a factor $2.4$, consistent at leading order with the predicted $1.85$. This channel is bounded by the conservation law -- the rate at which power can be deposited at wavenumber $k$ vanishes as $k^2$ -- and is hence confined to $k>k_c$, where the two initial conditions have already converged.

\subsection{Gravity in the conserved limit}
We now repeat the quench with the gravitational term active, remaining in the overdamped regime. As $k\rightarrow0$ the growth rate $L(k)$ tends to the constant $G$, the same for every mode, so gravity should multiply the whole large-scale spectrum by the single factor $e^{2G\tau}$, amplifying the memory without distorting its shape. With $G=0.01$, below $G^*$, and $\tau=120$ this is $e^{2G\tau}=11.0$, a prediction with no free parameters. Table~\ref{tab:amp} compares the $G=0.01$ ensemble against the $G=0$ ensemble at the same wavenumbers, pooled over three realisations at $L_{\rm box}=48\lambda_0$. The measured amplification at the largest resolved scale is $10.9\pm1.4$ (fig.~\ref{fig:inherit}, right). The exponent survives the amplification: the separation between the two fields is $0.94\pm0.62$, again consistent with $n_s$. Meanwhile the kinetically arrested state remains untouched: the position of the spectral peak, the spread of nearest-neighbour distances between density maxima, and the number of those maxima all match the $G=0$ ensemble to within $0.3\%$. Both instabilities of eq.~\ref{eq:free} thus operate in a single run, on separate scales: ordering builds the arrested state at $k_0$, while gravity amplifies the primordial memory as $k\rightarrow0$.

\begin{table*}[b]
\caption{Effect of the gravitational term in the conserved limit, $G=0.01$
versus $G=0$, pooled over three realisations at $L_{\rm box}=48\lambda_0$.
The amplification is measured at the largest resolved scale; the exponent
separation is $\alpha_{\rm primordial}-\alpha_{\rm white}$. Per-shell values
are shown in fig.~\ref{fig:inherit}, right.}
\label{tab:amp}
\begin{ruledtabular}
\begin{tabular}{lccc}
Quantity & $G=0$ & $G=0.01$ & Linear prediction \\ \colrule
Infrared amplification $S(k,\tau)/S_0(k)$ & --- & $10.9\pm1.4$ & $e^{2G\tau}=11.0$ \\
Exponent separation & $1.36\pm0.50$ & $0.94\pm0.62$ & $n_s=0.9665$ \\
Spectral peak position & reference & within $0.3\%$ & unchanged \\
Nearest-neighbour spread (CV) & reference & within $0.3\%$ & unchanged \\
Number of density maxima & reference & within $0.3\%$ & unchanged \\
\end{tabular}
\end{ruledtabular}
\end{table*}

\subsection{Real-space morphology}
\begin{figure}[t]
\includegraphics[width=\columnwidth]{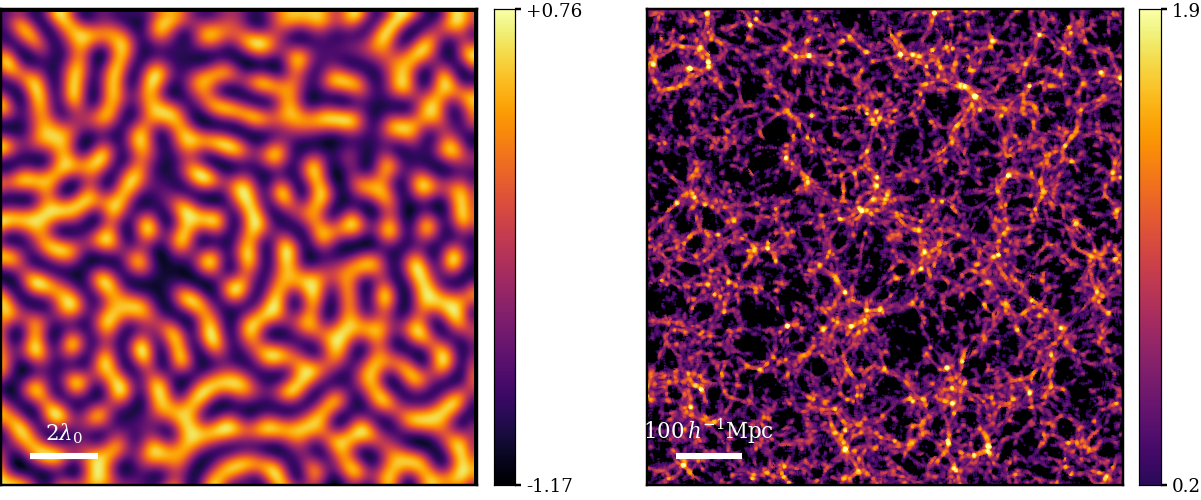}
\caption{The two kinetic limits of eq.~\ref{eq:free} from initial fields with
primordial infrared spectrum $S_0\propto k^{n_s}$. Left: the kinetically
arrested state from the conserved overdamped limit, eq.~\ref{eq:cons1}, at
$\tau=120$, on a slice of $0.18\lambda_0$ thickness. Right: the cosmic web
from the inertial limit with Hubble drag, eq.~\ref{eq:cons}, under the full
functional at $r=-1$, $k_0=4\,h\,$Mpc$^{-1}$, at $a=1$, on an $8\,h^{-1}$Mpc
slab. Both panels show one quarter of the simulation box.}
\label{fig:real}
\end{figure}

The structure factor does not uniquely determine the underlying morphology, so we characterise each end state in real space with three complementary diagnostics. The coefficient of variation CV of the distance between neighbouring density maxima vanishes for a crystal and takes the value $[\Gamma(5/3)/\Gamma(4/3)^2-1]^{1/2}=0.363$ for a three-dimensional Poisson process~\footnote{As follows from the nearest-neighbour distribution $p(r)=3\lambda r^2e^{-\lambda r^3}$, $ \lambda=4\pi\rho/3$.}; our simulations yield ${\rm CV}=0.25$-$0.27$, a disordered system distinctly more even than random. The skewness of the density field, which distinguishes isolated maxima from connected ridges, is $+0.19$ in all runs, and the number of density maxima scales with system volume to within $0.2\%$ over a factor of $512$, confirming a well-defined bulk state rather than a finite-size morphology. Together these measures identify the large-system state as a spatially homogeneous labyrinth which, with an exponent $\alpha=0.93\pm0.20$, is a class III DHU (fig.~\ref{fig:real}, left), and corresponds to the DHU arrangements that regulate density fluctuations in several systems at multiple scales~\cite{merminod2015transition,ma2017random,huang2021circular,leoni2025confinement}. At the same quench parameters, smaller systems form droplet morphologies, as expected~\cite{qian2003defect,bordeu2015localized,ainsworth2020fractional}.

% =====================================================================
\section{The inertial limit: the cosmic web}
\label{sec:cosmo}
% =====================================================================

\subsection{The cosmic web under the full functional}
\label{sec:web}
We next evolve the primordial field in the inertial limit of eq.~\ref{eq:cons} with the full functional, at the pressureless point $r=-1$ and $\lambda_0=1.57\,h^{-1}$Mpc, an order of magnitude below the structure-forming scales. We simulate $512$ million particles on an $800^3$ mesh in a $300\,h^{-1}$Mpc box, the mesh being required by eq.~\ref{eq:resbound} at this $k_0$ and box. From the identical initial field, on the identical mesh, we also evolve the $\lambda_0\rightarrow0$ limit. Therefore, the two runs differ only in the ordering term.

The field develops a cosmic web of filaments and dense nodes separated by extended voids (fig.~\ref{fig:real}, right). At $R_s=2\,h^{-1}$Mpc the full functional gives volume fractions of $0.404$ voids, $0.428$ sheets, $0.152$ filaments and $0.016$ nodes; gravity alone, from the same initial field, gives $0.370/0.445/0.173/0.012$. The full functional therefore reproduces the gravitational tidal skeleton to within $0.03$ in every class, with slightly more void and fewer filaments. A $150\,h^{-1}$Mpc box at the same cell size gives $0.401/0.425/0.156/0.018$, so the classification is converged with respect to box size. The density contrast, however, saturates at $\delta_{\max}=19.9$, with $\delta=(\rho-\bar\rho)/\bar\rho$, against $3.8\times10^3$ under gravity: the quartic term that arrests the labyrinth in the conserved limit also caps compression here.

\subsection{The ordering-to-gravity crossover}
\label{sec:cross}
The inertial sector approaches the gravitational limit under two independent conditions, established by the runs of Table~\ref{tab:cross}.\newline 
The medium must be pressureless. Away from $r=-1$ the non-gravitational part of the kernel, $r+(1-u^2)^2$, does not vanish as $u\rightarrow0$ and acts as an inverse compressibility. At $r=-0.3$, $G=0.030$ the bracket of eq.~\ref{eq:kernel} changes sign and becomes positive, resisting compression, over $0.11<k<0.31\,h\,$Mpc$^{-1}$, i.e., $20<\lambda<56\,h^{-1}$Mpc, which corresponds to the band in which sheets and filaments form. Collapse is suppressed and the box remains $98.8\%$ void and $\delta_{\max}$ saturates at $4.8$. Deepening the quench to $r=-0.4$ and raising $G$ to $0.050$ closes the repulsive band but leaves the force between $20\%$ and $60\%$ of Newtonian across the same range, and the morphology is essentially unchanged ($92.7\%$ void, $\delta_{\max}=5.3$). Only at $r=-1$, where the constant term cancels identically and the residual is $O(u^4)$, is the kernel attractive at every scale below $k_0$.

At $r=-1$ the morphology is then set by $\lambda_0$ alone. With $\lambda_0=3.1\,h^{-1}$Mpc the gravitational skeleton is recovered but the filaments remain diffuse; with $\lambda_0=1.57\,h^{-1}$Mpc the T-web fractions agree with gravity as reported above. The two sectors of eq.~\ref{eq:free} are therefore continuously connected, with $r$ and $\lambda_0$ selecting between them.

%The density contrast does not converge: it remains capped in every
%full-functional run, $\delta_{\max}=19.9$ at $\lambda_0=1.57\,h^{-1}$Mpc.
%Since the cubic coefficient in $\mu$ scales as $1/(Gk_0^2)$ and $G$ is
%bounded, the saturation contrast grows only as $k_0$; recovering the
%gravitational value would require $\lambda_0$ smaller by a further order of
%magnitude. The functional builds the gravitational tidal skeleton and fills it
%with soft matter.

\begin{table*}[b]
\caption{Crossover of the inertial limit towards gravity. T-web volume
fractions at $a=1$ with $\lambda_{\rm th}=0.2$ and $R_s=2\,h^{-1}$Mpc, except
for the $\lambda_0\to0$, $1200\,h^{-1}$Mpc entry, which uses
$R_s=6\,h^{-1}$Mpc set by its coarser mesh and is not directly comparable. The
two ``paired'' rows share the initial field and the $800^3$ mesh.}
\label{tab:cross}
\begin{ruledtabular}
\begin{tabular}{lcccccccc}
$r$ & $k_0$ & $\lambda_0$ & $L_{\rm box}$ & void & sheet & filament & node & $\delta_{\max}$ \\
 & [$h\,$Mpc$^{-1}$] & [$h^{-1}$Mpc] & [$h^{-1}$Mpc] & & & & & \\ \colrule
$-0.3$ & 0.5 & 12.6 & 1200 & 0.988 & 0.012 & 0.000 & 0.000 & 4.8 \\
$-0.4$ & 0.5 & 12.6 & 1200 & 0.927 & 0.070 & 0.002 & 0.000 & 5.3 \\
$-1$   & 2.0 & 3.14 & 300  & 0.445 & 0.414 & 0.131 & 0.010 & 12.8 \\
$-1$   & 4.0 & 1.57 & 150  & 0.401 & 0.425 & 0.156 & 0.018 & 16.5 \\
$-1$, paired & 4.0 & 1.57 & 300 & 0.404 & 0.428 & 0.152 & 0.016 & 19.9 \\
$\lambda_0\to0$, paired & --- & --- & 300 & 0.370 & 0.445 & 0.173 & 0.012 & 3790 \\
$\lambda_0\to0$, $128^3$ & --- & --- & 300 & 0.396 & 0.436 & 0.157 & 0.011 & 311 \\
$\lambda_0\to0$ & --- & --- & 1200 & 0.482 & 0.383 & 0.124 & 0.011 & 283--414 \\
\end{tabular}
\end{ruledtabular}
\end{table*}

\subsection{The infrared under the full functional}
\label{sec:pair}
The same pair of runs tests whether the ordering term alters the growth of the large-scale spectrum. Throughout this section the infrared statistic is $S_{\rm IR}/D^2$: $S_{\rm IR}$ is the mean of $S(k)$ over a band $k<mk_{\min}$, with $k_{\min}=2\pi/L_{\rm box}$ and each shell weighted by its number of Fourier modes, and $D(a)$ is the linear growth factor~\footnote{$D(a)$ is the solution of $\ddot\delta+2H\dot\delta=4\pi G_N\bar\rho_m\delta$ normalised to $D(a_0)=1$; linear evolution scales every mode of $S(k)$ by $D^2(a)$.}, normalised to the initial value. Since $D(a)$ describes the growth of density perturbations while they remain small, a flat $S_{\rm IR}/D^2(a)$ means that the infrared power grows as linear theory predicts.

\begin{table*}[b]
\caption{Paired test of the ordering term. $[S(k,a=1)/S(k,a_0)]/D^2$
averaged over each band with mode-count weights, in the $300\,h^{-1}$Mpc box,
$k_{\min}=0.0209\,h\,$Mpc$^{-1}$. ``PFC on'' and ``PFC off'' share the initial
field and the $800^3$ mesh; the last column is the mean and standard
deviation over four realisations of gravity alone on a $128^3$ mesh. The
$k_0=2$ rows are the $400^3$ run of Table~\ref{tab:cross} in the same box,
compared with the $128^3$ mean.}
\label{tab:pair}
\begin{ruledtabular}
\begin{tabular}{lccccc}
band & $\lambda$ [$h^{-1}$Mpc] & PFC on & PFC off & on/off & gravity, $128^3$ \\ \colrule
$k<2k_{\min}$ & $>150$ & 0.913 & 0.959 & 0.952 & $0.972\pm0.035$ \\
$k<3k_{\min}$ & $>100$ & 0.941 & 1.014 & 0.927 & $0.976\pm0.028$ \\
$k<4k_{\min}$ & $>75$  & 0.870 & 0.949 & 0.917 & $0.955\pm0.016$ \\
$k<6k_{\min}$ & $>50$  & 0.846 & 0.988 & 0.856 & $0.944\pm0.033$ \\ \colrule
$k_0=2$, $k<3k_{\min}$ & $>100$ & 0.855 & --- & $0.876^{\rm a}$ & \\
$k_0=2$, $k<6k_{\min}$ & $>50$  & 0.702 & --- & $0.744^{\rm a}$ & \\
\end{tabular}
\end{ruledtabular}
\raggedright\footnotesize $^{\rm a}$Relative to the $128^3$ gravity mean.
\end{table*}

\begin{figure}[t]
\includegraphics[width=\columnwidth]{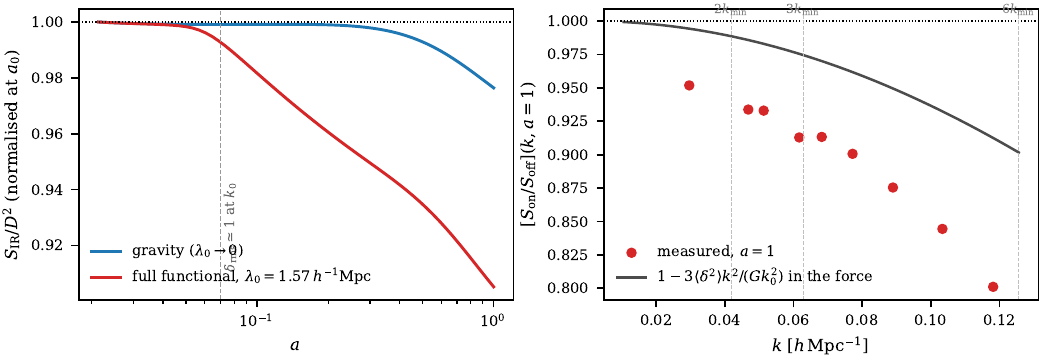}
\caption{Paired test of the ordering term on the infrared. Left: $S_{\rm IR}/D^2$ over $k<4k_{\min}$ against $a$ for the two runs of Sec.~\ref{sec:web}, each normalised to its first output; same initial field, same $800^3$ mesh. The dashed line marks the saturation of the $k_0$ modulation. Right: the ratio of the two runs shell by shell at $a=1$ (points) and the suppression produced by the mean-field pressure of eq.~\ref{eq:eps} in the linear growth equation (curve), with $\langle\delta^2\rangle(a)$ taken from the full-functional run.}
\label{fig:pair}
\end{figure}

Over the band $k<3k_{\min}=0.021$--$0.063,h,$Mpc$^{-1}$ ($\lambda=100$--$300,h^{-1}$Mpc, the band that carries most of the variance on the scale $R_H\simeq70,h^{-1}$Mpc at which the galaxy distribution is measured to approach homogeneity~\cite{hogg2005cosmic,sarkar2009scale,scrimgeour2012wigglez,dias2023probing}), the ratio at $a=1$ is $0.941$ under the full functional against $1.014$ under gravity alone (Table~\ref{tab:soft}, fig.~\ref{fig:pair}). The infrared power therefore still tracks the linear growth factor to $94\%$ while $D^2$ grows by a factor of $1363$, but measurably below pure gravity: the ratio of the two, $0.927$, lies outside the realisation scatter of the band, $\pm0.028$, measured over four seeds of gravity alone.

The deficit is smooth and monotonic in scale. Shell by shell, the ratio of the two runs is $0.952$, $0.934$, $0.933$, $0.913$, $0.913$, $0.901$, $0.875$, $0.845$ and $0.801$ at $k=0.030$, $0.047$, $0.051$, $0.062$, $0.068$, $0.077$, $0.089$, $0.103$ and $0.118,h,$Mpc$^{-1}$ (fig.~\ref{fig:pair} right panel). The value in the lowest shell, $\lambda=212,h^{-1}$Mpc, is comparable to the $\pm3.5\%$ realisation scatter of that eight-mode shell and is not individually significant, while the band averages are. The deficit also accumulates steadily in time. In the unweighted estimator logged during the runs, the two coincide at $a=0.038$ and their ratio falls to $0.993$, $0.972$, $0.954$, $0.941$ and $0.927$ at $a=0.074$, $0.14$, $0.27$, $0.52$ and $1$: a growth-rate deficit of $\sim3\%$ per $e$-fold of $a$, switched on at $a\simeq0.07$ (fig.~\ref{fig:pair} left panel). This is the epoch at which the ordering modulation saturates: $\delta_{\rm rms}$ in the full-functional run is $0.37$ at $a=0.038$ and $0.79$ at $a=0.074$, against $0.19$ and $0.35$ under gravity, and stays near unity thereafter. Finally, the deficit increases with $\lambda_0$: the $k_0=2,h,$Mpc$^{-1}$ run in the same box shows $1.6$-$1.7$ times the deficit in every band, $12\%$ over $k<3k_{\min}$ and $24\%$ over $k<6k_{\min}$.

The linear force of the ordering term is irrelevant at these scales: relative to gravity it is $(2u^4-u^6)/G$ with $u=k/k_0\lesssim0.03$, below $10^{-7}$. Once the modulation has saturated, the quartic term becomes relevant. Writing $n=n_L+n_0$, with $n_0$ the saturated $k_0$-scale texture, the low-$k$ part of $(n^3)_k$ is $3\langle n_0^2\rangle n_{L,k}$ at leading order, so the chemical potential of a large-scale mode becomes $\mu_k=[r_{\rm eff}+(1-u^2)^2]n_k-Gn_k/u^2$ with
\begin{equation}
    r_{\rm eff}=r+3\langle\delta^2\rangle ,
    \label{eq:reff}
\end{equation}
the same shift that the mean offset $\bar n$ produces in the conserved limit (Sec.~\ref{sec:soft}), now generated by the fluctuations. At $r=-1$ and $\langle\delta^2\rangle\simeq1$ this is $r_{\rm eff}\simeq+2$, an inverse compressibility: the medium is no longer pressureless. In the kernel eq.~\ref{eq:kernel} it multiplies the gravitational force on a large-scale mode by
\begin{equation}
    1-\epsilon(k,a),\qquad
    \epsilon=\frac{3\langle\delta^2\rangle(a)\,k^2}{G\,k_0^2}.
    \label{eq:eps}
\end{equation}
Integrating the linear growth equation with this force, using $\langle\delta^2\rangle(a)$ from the run's own $\delta_{\rm rms}$, gives a suppression of $S/D^2$ at $a=1$ of $0.994$, $0.976$, $0.962$ and $0.913$ at $k=0.030$, $0.062$, $0.077$ and $0.118\,h\,$Mpc$^{-1}$: the measured scale dependence, at a quarter to a half of the measured magnitude (fig.~\ref{fig:pair}b). The shortfall is expected. The coefficient $3\langle\delta^2\rangle$ is the Gaussian value of $\partial\langle n^3\rangle/\partial n_L$, and the saturated texture is strongly non-Gaussian, with $\delta$ up to $20$ in the ridges; a mean-field estimate fixes the form and the order of magnitude, not the coefficient. Equation~\ref{eq:eps} does fix the two scalings that matter: $\epsilon\propto k^2$, so the suppression is an envelope on the spectrum rather than a reshaping of it, and $\epsilon\propto\lambda_0^2/G$, so it vanishes as $\lambda_0\rightarrow0$. The pressureless point $r=-1$ is therefore exact only at the linear level. Nonlinear saturation of the ordering modulation endows the full functional with a residual pressure $\propto\langle\delta^2\rangle\lambda_0^2/G$, and the cold-matter limit of eq.~\ref{eq:free} is $\lambda_0\rightarrow0$.

\subsection{The infrared in the cold-matter limit}
\label{sec:ir}
Testing infrared preservation on genuinely hyperuniform scales requires wavenumbers below the turnover of $P(k)$ at $k_{eq}=0.0158\,h\,$Mpc$^{-1}$, hence a box larger than $10^3\,h^{-1}$Mpc. Resolving $\lambda_0$ in such a box is not possible: reproducing the gravitational morphology requires $k_0\gtrsim4\,h\,$Mpc$^{-1}$ (Table~\ref{tab:cross}), and with $L_{\rm box}\gtrsim1200\,h^{-1}$Mpc eq.~\ref{eq:resbound} demands $N_{\rm mesh}>2.07\times4\times1200/\pi\simeq3.2\times10^3$, a $3200^3$ mesh, some $500$ times the cost of the runs reported here and of order $10^3$~GB of memory. Moreover, any finite $\lambda_0$ carries the residual pressure of Sec.~\ref{sec:pair}, which would have to be corrected for. We therefore perform this test in the cold-matter limit $\lambda_0\rightarrow0$ of eq.~\ref{eq:free}, which Sec.~\ref{sec:cross} shows to be approached continuously from the full functional. We simulate $64$ million particles on a $400^3$ mesh in a $1200\,h^{-1}$Mpc box, three independent realisations, and measure the band $k<3k_{\min}=0.0052$--$0.016\,h\,$Mpc$^{-1}$, wavelengths from $1200$ down to $400\,h^{-1}$Mpc.

\begin{figure}[t]
\includegraphics[width=\columnwidth]{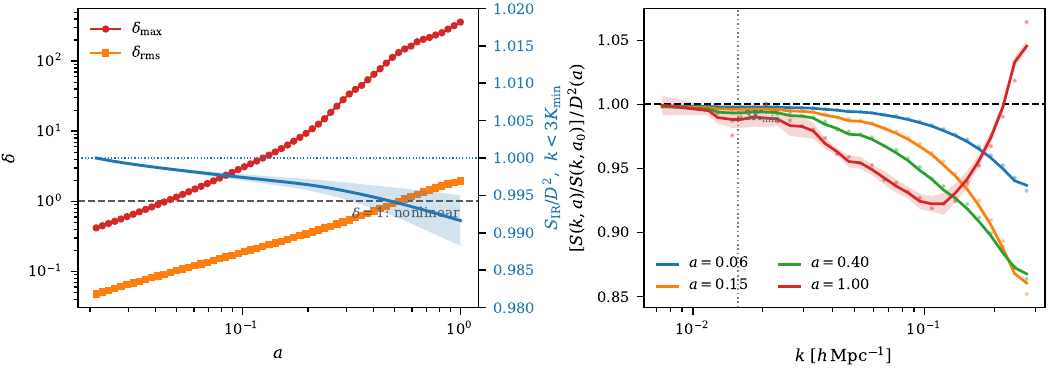}
\caption{The cold-matter limit $\lambda_0\to0$ of eq.~\ref{eq:free}: three realisations in a $1200\,h^{-1}$Mpc box. Left: growth of $\delta_{\max}$ (red circles) and $\delta_{\rm rms}$ (orange squares); the black dashed line marks $\delta=1$, the onset of nonlinearity. The blue curve (right axis) is $S_{\rm IR}/D^2(a)$ over $k<3k_{\min}$, normalised to its initial value; the shaded band is the standard deviation over the three realisations. Right: $[S(k,a)/S(k,a_0)]/D^2(a)$ at four epochs, averaged over the same three realisations ($a=0.06$ blue, $0.15$ orange, $0.40$ green, $1.00$ red). Faint points are the seed-averaged individual shells; solid curves are running means over $0.1$ decades in $k$, weighted by mode count. The shaded band around the $a=1$ curve is the standard deviation over realisations, shown for that epoch alone for clarity. Scale-independent growth appears as unity (dashed line).}
\label{fig:cosmo}
\end{figure}

$\delta_{\max}$ and $\delta_{\rm rms}$ cross the nonlinearity threshold $\delta=1$ at different times (fig.~\ref{fig:cosmo}, left): $\delta_{\max}$ while the field is still globally linear, $\delta_{\rm rms}$ only near the end of the run, the signature of hierarchical collapse, and the reason the infrared result is not trivial: the largest scales are preserved while collapsed objects are already present. At $a=1$ the field is strongly nonlinear, with $\delta_{\rm rms}=1.96$ and $\delta_{\max}=283$--$414$ across the three realisations.

The blue curve in fig.~\ref{fig:cosmo}, left, is $S_{\rm IR}/D^2(a)$; the shaded band is the standard deviation over the three realisations. At $a=1$ the ratio is $0.992\pm0.003$: the infrared power tracks $D^2(a)$ to better than one per cent while $D^2$ grows by a factor of $1363$ and the field itself becomes strongly nonlinear. The small residual deficit is not located at the fundamental modes: restricting the average to $k<2k_{\min}$ gives $0.9992\pm0.0076$, and widening it to $k<6k_{\min}$ gives $0.9883\pm0.0033$. The departure from unity grows with the upper edge of the band, as expected if it originates in the quasi-linear modes rather than in the largest scales.

Figure~\ref{fig:cosmo}, right, resolves the same comparison by wavenumber. Below $3k_{\min}$ the four epochs lie together on the dashed line: averaged over the three longest-wavelength shells the ratio at $a=1$ is $0.996\pm0.002$. Gravity rescales the large-scale spectrum without redistributing power across it. Above $k\simeq0.05\,h\,$Mpc$^{-1}$ the epochs separate, and by $a=1$ the ratio exceeds unity for $k\gtrsim0.2\,h\,$Mpc$^{-1}$. That excess is the transfer of power to small scales which accompanies nonlinear collapse, and it is the counterpart of the preservation seen in the infrared: the power that leaves the quasi-linear scales does not come from the largest ones.

\subsection{Local spectral slope and the galaxy distribution}
\label{sec:alpha}
\begin{figure}[t]
\includegraphics[width=\columnwidth]{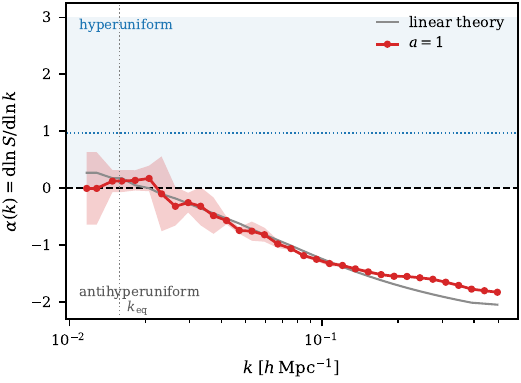}
\caption{Local spectral slope $\alpha(k)=d\ln S/d\ln k$ at $a=1$ in the $\lambda_0\to0$ runs, mean and standard deviation over three realisations (points), against the slope of linear theory (line), which is also that of the initial condition. The vertical line marks $k_{eq}$.}
\label{fig:alpha}
\end{figure}

Hyperuniformity is a statement about the sign of the local spectral slope $\alpha(k)\equiv d\ln S/d\ln k$ as $k\rightarrow0$: $\alpha>0$ denotes hyperuniformity, $\alpha=0$ a Poisson point pattern, $\alpha<0$ antihyperuniformity. We measure $\alpha$ in the $\lambda_0\to0$ runs by a mode-count-weighted least-squares fit of $\log S$ against $\log k$ in a sliding window of $0.15$ decades, after subtracting the Poisson term from the particle structure factor, since $\alpha$ characterises the underlying continuous field rather than its discrete sampling (fig.~\ref{fig:alpha}).

Averaged over the three realisations, the measured slope agrees with linear theory to $|\Delta\alpha|\le0.08$ over $0.04<k<0.14\,h\,$Mpc$^{-1}$ and is shallower by up to $0.27$ above $k\simeq0.15\,h\,$Mpc$^{-1}$, where nonlinear collapse transfers power to small scales -- the same effect that raises the $a=1$ curve of fig.~\ref{fig:cosmo}, right, above unity. Since $D^2(a)$ does not alter $d\ln S/d\ln k$, the linear-theory curve is also the slope of the initial condition; the agreement at low $k$ is a second, independent statement that nonlinear evolution does not reshape the infrared spectrum.

The field is accordingly hyperuniform on the largest scales and increasingly antihyperuniform below $\lambda\simeq300\,h^{-1}$Mpc, reproducing the crossover reported for the observed galaxy distribution~\cite{philcox2023disordered}; here it arises from the functional rather than being assumed. Two limits apply to the comparison. The sign change occurs between $k=0.021$ and $0.026\,h\,$Mpc$^{-1}$ because the transfer function places the turnover of $P(k)$ there, and what the measurement shows is that nonlinear evolution leaves it in place. And the positive branch is not individually resolved: $\alpha=+0.17\pm0.22$ at $k=0.021\,h\,$Mpc$^{-1}$, because the number of Fourier modes below $k_{eq}$ in a $1200\,h^{-1}$Mpc box is small, a limitation of cosmic variance rather than of the model.

\section{Discussion and conclusions}
\label{sec:conc}
We have shown that a single density-functional free energy can support two apparently disparate forms of disordered structure: a DHU bicontinuous labyrinth phase embracing multiple systems and length-scales, and a gravitationally generated cosmic web. The two sectors are continuously connected, with $r$ and $\lambda_0$ selecting between them, and their distance is measurable. The pressureless point $r=-1$ is exact at the linear level; once the ordering modulation saturates, its mean-field back-reaction $r_{\rm eff}=r+3\langle\delta^2\rangle$ leaves the medium with a residual pressure $\propto\langle\delta^2\rangle\lambda_0^2/G$ that suppresses the growth of the largest scales by a smooth $k^2$ envelope, $7\%$ over $\lambda=100$--$300\,h^{-1}$Mpc at $\lambda_0=1.57\,h^{-1}$Mpc, and vanishes in the cold-matter limit $\lambda_0\rightarrow0$. The tidal skeleton of the web is reproduced by gravity to within $0.03$ across all classes, with the sole exception of the density contrast in collapsed regions, where deviations arise from the saturation imposed by the quartic term.

In both kinetic limits the infrared spectrum is not produced by the nonlinear dynamics. Conservation laws restrict the long-wavelength contribution generated during evolution to subleading terms, while the initial spectrum remains dominant. For $S_0(k)\propto k^{n_s}$ with $n_s<4$, this yields a robust infrared memory that is frozen by kinetic arrest and amplified by gravity without distortion in the cold-matter limit, where the spectrum tracks $D^2$ to better than one per cent over a factor $1363$ of growth, and with a $k^2$ envelope, not a reshaping, when the ordering term is active. These results indicate that suppressed long-wavelength fluctuations are not necessarily diagnostic of an equilibrium state or a specific ordering mechanism, but can instead emerge as a consequence of the dynamical inheritance of the initial conditions. More broadly, the model provides a common framework for studying infrared memory across microscopic arrested matter and gravitational structure formation.

\begin{acknowledgments}
The author is truly grateful to Fabio Leoni and Andrea Gabrielli for insightful discussions. The author conceived the study, developed the mathematical framework, and wrote the code to simulate the dynamics and generate the underlying data. Claude Opus 5 helped generate codes to prepare plots from the data the author produced. The author reviewed and verified all codes and resulting figures.
\end{acknowledgments}

\bibliography{apssamp}
\end{document}